\documentclass{elsarticle}

\usepackage{color,amsmath,amssymb}

\newcommand{\3}{$_{3}$}

\newcommand{\cm}{cm$^{-1}$}

\newcommand{\lnr}{^{\ell}}

\newcommand{\abinitio}{\textit{ab initio}}

\renewcommand{\thefootnote}{\fnsymbol{footnote}}

\begin{document}

\title{A Computed Room Temperature Line List for Phosphine}

\author{Clara Sousa-Silva*$^a$}

\author{Sergei~N.~Yurchenko$^a$}

\author{Jonathan Tennyson$^a$}

\address{$^a$Department of Physics and Astronomy, University College London,\\
London, WC1E 6BT, UK}

\date{\today}

\begin{abstract}

  An accurate and comprehensive room temperature rotation-vibration
  transition line list for phosphine ($^{31}$PH$_{3}$) is computed
  using a newly refined potential energy surface and a previously
  constructed {\it{ab initio}} electric dipole moment surface. Energy
  levels, Einstein A coefficients and transition intensities are
  computed using these surfaces and a variational approach to the
  nuclear motion problem as implemented in the program TROVE. A 
  ro-vibrational spectrum is computed, covering the wavenumber range $0$ to $8000$ cm$^{-1}$.
   The resulting line list, which is appropriate for temperatures up to 300 K, consists of a total of
  $137$ million transitions between $5.6$ million energy levels.  Several of the band
  centres are shifted to better match experimental transition
  frequencies.  The line list is compared to the most recent HITRAN
  database and other laboratorial sources.  Transition wavelengths and
  intensities are generally found to be in good agreement with the
  existing experimental data, with particularly close agreement for the
  rotational spectrum. An analysis of the comparison between the
  theoretical data created and the existing experimental data is
  performed, and suggestions for future improvements and assignments
  to the HITRAN database are made.

\end{abstract}

\maketitle

\newpage

\section{Introduction}
\label{sec:intro}

Phosphine (PH$_{3}$) is an extremely toxic, rigid molecule found in the lower troposphere of the Earth and is expected to be of great importance for the spectroscopic studies of giant-planets, particularly for probing the deeper layers of the atmospheres \cite{92TaLaLe}, and cool stars. Phosphine has for a long time known to be abundant in Jupiter \cite{09FlOrTe,75PrLeXX,98EdAtTr}, where it is partially responsible for the continuum opacity in the $5$ $\mu$m region of the atmosphere \cite{79BeTaXX} and where it is a marker for vertical convection zones. It has also been found on Saturn \cite{92TaLaLe,09FlOrTe}, and is expected to be present in extrasolar gas giants.

Further spectroscopic investigations of phosphine in the atmospheres of astrophysical bodies requires an extrapolation of the spectra over a wide range of temperatures, pressures and path lengths. The ability to accurately identify and interpret phosphine features in astrophysical spectra is dependent on the existence of a complete description of the phosphine spectra, for all relevant temperatures.

The current  CDMS \cite{05MuScSt} and HITRAN \cite{jt453} databases encapsulate decades of accurate laboratorial measurements of phosphine spectra at room temperature. However, between the two databases, only a total of 22230 lines are described, all within the $0-3600$ cm$^{-1}$ region. CDMS contains transitions between $0-300$ cm$^{-1}$ with J values of up to 34, sensitive to 10$^{-36}$ cm/molecule, while HITRAN has transitions between $770-3600$ cm$^{-1}$ stronger than 10$^{-28}$ cm/molecule with $J\leq23$. These databases are designed to be valid for temperatures below 300 K. Other sources (see Table \ref{sources} below) mostly overlap with the HITRAN and CDMS databases, and only add little more to these compilations.

The quantity and variety of spectral information required to correctly characterise hot astrophysical bodies is often beyond what can be expected to be delivered experimentally. The ExoMol project  \cite{jt528} (see {\it{www.exomol.com}}) aims to provide a solution to this problem by performing accurate quantum mechanical calculations, with the aim of providing appropriate spectroscopic data on all relevant molecules for the analysis and characterisation of cool stellar and exoplanet atmospheres. This is, however, a challenging alternative to measuring spectra, requiring highly accurate potential energy surfaces (PES), dipole moment surfaces (DMS) and an ability to generate precise wave functions and eigenvalues for the nuclear Schr\"odinger equation and the associated transition intensities. In practice, at least for the PESs, it is also necessary to incorporate a certain amount of experimental refinement.

Our aim is to develop complete line lists which could be used to accurately model the atmospheres of stars and Jupiter-like planets, for which a high temperature ($\geq1500$ K) line list for PH$_{3}$ is of special importance. In order to establish and test the production of such a line list, which will consist of approximately 1 billion transitions, it is first pertinent to perform a simulation of the room-temperature intensities for this molecule, which can be tested against existing reliable experimental sources. In the present paper a cool ($300$ K) line list  is presented.

There have been several previous attempts at {\it{ab initio}} studies of the behavior and characteristics of phosphine spectra. Wang et al \cite{00WaShZh} computed a PES using coupled cluster CCSD(T) theory and Dunning's correlation consistent cc-pVTZ basis sets, followed by a vibrational analysis based on second-order perturbation theory. They determined several spectroscopic constants and fundamentals largely within $4$ cm$^{-1}$ of the observed results.
Zheng et al \cite{01HeZhHu} calculated an {\it{ab initio}} three-dimensional P-H stretching DMS using density functional theory and found absolute band intensities agreed with observed ones within a factor of 2.
Yurchenko et al \cite{03YuCaJe} presented a calculated, albeit empirically refined, PES and, using a variational method, calculated the rotational energy levels in the vibrational ground state of PH$_{3}$ for $J\leq80$ \cite{05YuThPa}, thus establishing for the first time the existence of six-fold near-degenerate ro-vibrational energy clusters for this molecule. Subsequently, Yurchenko et al \cite{06YuCaTh} computed an entirely {\it{ab initio}} six-dimensional dipole moment surface (DMS) at the CCSD(T)/aug-cc-pVTZ level of theory for the electronic ground state of PH$_{3}$. This is the DMS that is used in the present work, and is described in detail below. It gives calculated transition moments within an average of $22.5\%$ of the experimental data. The same paper also presented a PES determined by empirically refining an existing {\it{ab initio}} surface.
Ovsyannikov et al \cite{08OvThYu2} complemented the variationally calculated PES and DMS of Yurchenko et al \cite{06YuCaTh} by calculating a PES at the CCSD(T) level using aug-cc-pV(Q+d)Z and aug-cc-pVQZ basis sets for P and H, respectively, and then presenting a list of computed vibrational transition moments for the electronic ground state of PH$_{3}$ \cite{08OvThYu}. The PES used here is a refined version of the PES presented by Ovsyannikov et al \cite{08OvThYu2}, and is discussed further below.
Recently, Nikitin et al produced a new PES \cite{09NiHoTy} and used it to calculate several vibrational energy levels with only a marginal deviation from experiment. Nikitin et al then successfully modelled the lower three polyads of phosphine using a variational approach \cite{09NiChBu}. The resulting line list is discussed and compared to the present work in Section {\bf{4.6}}.
The phosphine line list presented here can be considered the analogue of the ammonia line lists computed by Yurchenko et al \cite{jt466}, using TROVE.

\section{Background}
Phosphine is a well behaved symmetric top molecule belonging to the group-15 hydrides. The phosphorous atom is positioned on the axis of symmetry, perpendicular to the plane containing the equidistant three hydrogen atoms. As is common with molecules having $C_{\rm 3v}$(M) symmetry \cite{BJ}, there is a splitting of rotational levels with $K=3n$ (where $n\geq1$) in non degenerate vibrational states \cite{83CaLoTr}.
Phosphorous has only one non-synthetic, stable isotope, $^{31}$P, so only $^{31}$PH$_{3}$ is considered in the present work.

Phosphine has near degenerate $\nu_{1}$ symmetric and $\nu_{3}$ antisymmetric stretching modes \cite{00WaShZh} with frequencies of about twice that of the asymmetric bending mode, $\nu_{4}$. It is therefore natural to consider its spectrum in terms of polyads, which will be described in section 3.1. There is also a strong Coriolis interaction between the $\nu_{2}$ and $\nu_{4}$ bending bands which causes large distortion effects in observations \cite{81TaDaGo}.
Recorded phosphine spectra show no inversion splitting. The tunnelling effect found in ammonia is expected to occur in phosphine but so far attempts to detect it have failed due to its considerably higher potential energy barrier ($12300$ cm$^{-1}$) between the two symmetry-related minima \cite{81BeBuPo,92ScLaPy}.

In the absence of the inversion splitting, phosphine is characterized by the $C_{\rm 3v}$(M) molecular group symmetry, i.e. each eigenfunction transforms according to one of the irreducible representations $A_{1}$, $A_{2}$ and $E$, the latter of which is two-fold degenerate.

To fully describe the internal rovibrational motion of phosphine one needs the following minimal set of quantum numbers \cite{jt543}:
\begin{equation}
\label{e:quanta}
  n_1, n_2, n_3, n_4, {L_3}, {L_4}, L, \Gamma_{\rm vib}, J, K, \Gamma_{\rm rot}, \Gamma_{\rm tot},
\end{equation}
where  $L_3= |l_3|, L_4= |l_4|, L= |l|, K = |k| $. Here the vibrational quantum numbers $n_{1}$ (symmetric stretch), $n_{2}$ (symmetric bend), $n_{3}$ (asymmetric stretch) and $n_{4}$ (asymmetric bend) correspond to excitations of, respectively, the $\nu_{1}$, $\nu_{2}$, $\nu_{3}$ and $\nu_{4}$ modes. The doubly degenerate modes $\nu_{3}$ and $\nu_{4}$ require additional quantum numbers $L_{3} = |l_3|$ and $L_{4}=|l_4|$ describing the projections of the corresponding angular momenta (see, for example, Ref. \cite{BJ}).  The vibrational quantum number $L=|l|$, characterizes the coupling of $l_3$ and $l_4$. $\Gamma_{\rm rot}$, $\Gamma_{\rm vib}$,   and $\Gamma_{\rm tot}$ are, respectively, the symmetry species of the rotational, vibrational and total internal wave-functions in the molecular symmetry group $C_{\rm 3v}$(M), spanning $A_1$, $A_2$ and $E$. $J$ is the total angular momentum and $K=|k|, k= -J,\ldots ,J$ is the projection of the total angular momentum on the molecule fixed axis $z$.

Out of these twelve quantum numbers, only two are rigorously conserved quantum numbers ($J$ and $\Gamma_{\rm tot}$). This quantum number description is similar to that proposed by Down 
et al \cite{jt543} for ammonia.
The rigorous dipole selection rules include transitions between states with $\Gamma_{\rm tot}$ satisfying  $A_1 \leftrightarrows A_2$ and $E \leftrightarrows E$ and $\Delta{J}=0,\pm1$ only. Strong transitions obey the dipole selection rules $\Delta{J}=\pm1$ and $\Delta{K}=0$. There are also weakly allowed transitions which occur when, due to centrifugal distortion, the $C_{3v}$ geometrical symmetry of the molecule is broken and $K$ is no longer a good quantum number \cite{74ChOkXX}.

Only approximate selection rules can be associated with the vibrational quantum numbers $n_1$, $n_2$, $n_3$, $n_4$.

Table \ref{sources} presents an overview of the experimental spectra recorded for phosphine. This spans research starting from 1951 to the most recent body of work, published in 2009.
The naming convention for each individual source consist of two digits for the date and first two letters of up to four authors' surnames, e.g. S. Yurchenko, M. Carvajal, P. Jensen, H. Lin, J. Zheng \& W. Thiel 2005 becomes 05YuCaJeLi.

\newpage
\renewcommand{\thefootnote}{\fnsymbol{footnote}}
\newcommand{\specialcell}[2][c]{%
\begin{tabular}[#1]{@{}c@{}}#2\end{tabular}}

\begin{center}
\begin{table}[h!]
\small \tabcolsep=10pt
\renewcommand{\arraystretch}{1.4}
\caption{Experimental sources of phosphine spectra.\label{sources} }
\begin{tabular}{{l}{r}{c}{l}l}
\hline\hline
{\bf{Source}} & {\bf{No of lines}} &  {\bf{Range(cm$^{-1}$)}} & {\bf{Uncertainties(cm$^{-1}$)}} & {\bf{Intensities}}\\
\hline
71DaNeWoKl\cite{71DaNeWo}$^{a}$$^{c}$&$53$&$0.0-4.8$&Yes&No\\
77HeGo\cite{77HeGoXX}$^{a}$$^{c}$&$27$&$0.0-17.8$&No&Calculated\\
81BeBuPoSh\cite{81BeBuPo}$^{a}$$^{c}$&$68$&$0.0-35.6$&No&No\\
51LoSt\cite{51LoStXX} & $1$ &$0.94$  & $10^{-6}$ & No\\
81PiPoCo\cite{81PiPoCo}$^{c}$&$1$&$8.9$&$3.3\times10^{-7}$&No\\

69HeGo\cite{69HeGoXX}&$3$&8.9 - 17.8&$8.3\times10^{-6}$&No\\

06CaPu\cite{06CaPuXX}$^{a}$&$25$&8.9 - 35.6&$Yes$&No\\

81BeBuGeKr\cite{81BeBuGe}$^{a}$&$52$&10 - 35.7&$No$&No\\

74ChOk\cite{74ChOkXX}$^{a}$$^{c}$&$12$&$14.6-15.8$&$2.7\times10^{-5}$&Calculated\\

79KrMeSk\cite{79KrMeSk}$^{c}$&$4$&$17.1-17.8$&$1.7\times10^{-6}$&No\\

00FuLo\cite{00FuLoXX}&$155(+2500)$&$36-125(+750$-$1400)$&$0.004$&No\\

88FuCa\cite{88FuCaXX}$^{a}$&$118$&$44.5-166.6$&$0.002$&No\\

02BrSaKlCo\cite{02BrSaKl}$^{b}$&$\textgreater1060$&$770-1335$&$0.0002$&Some\\

81TaDaGo\cite{81TaDaGo}$^{b}$&$1318$&$818-1340$&$0.05$&Calculated\\

97AiHaSpKr\cite{97AiHaSp}&$62(+250^{d})$&$924-1085$&$6\times10^{-6}$&No\\

01HeZhHuLi\cite{01HeZhHu}&$28$&$992-9040$&$\textless0.01$&Yes\\

04SaArBoWa\cite{04SaArBo}&$26$&$995-1093$&No&Yes\\

02UlBeKoZh\cite{02UlBeKo}&$\geq4500$&$1750-9200$&$\leq0.0005$&No\\

92TaLaLeGu\cite{92TaLaLe}$^{b}$&$6605$&$1885-2445$&$0.0054$&Yes\\

05WaChChDi\cite{05WaChCh}&$1760$&$\specialcell{1950 - 2480\\3280 - 3580}$&\textless6\%&Yes\\

80BaMaNaTa\cite{80BaMaNa}&$1244$&$2184-2446$&$0.005$&Approximate\\

02Suarez\cite{02SuXXXX}&$138$&$2300-2381$&0.006&Yes\\

06BuSaKlBr\cite{06BuSaKl}$^{b}$&$8075$&$2721-3601$&$9.2\times10^{-4}$&Yes\\

07Kshiraga\cite{07KsXXXX}&$400$&$2730 - 3100$&$3.9\times10^{-4}$&No\\

73MaSaOl\cite{73MaSaOl}&$414$&$2760-3050$&No&No\\

04UlBeKoZh\cite{04UlBeKo}&$\textgreater700$&$3280-3580$&0.005&Transmittance\\
\hline\hline

\end{tabular}
\end{table}
\end{center}

{\footnotesize{
\flushleft{
$^{a}$Used in CDMS \cite{05MuScSt}.\\
$^{b}$Used in HITRAN 08 \cite{jt453}.\\
$^{c}$Used in JPL\cite{98PiPoCo}.\\
$^{d}$Compiled from other sources.\\
}}}

\vspace{2cm}

\section{Method}

The production of the phosphine line list presented here relies on the computational program TROVE (Theoretical ROVibrational Energies)  \cite{07YuThJe}, which uses a general variational approach to calculate the ro-vibrational energies for small semi-rigid polyatomic molecules of arbitrary structure. It employs a numerical finite basis representation.
The nuclear kinetic energy operator is numerically constructed through a recursive scheme using a Taylor series expansion in terms of the molecule's internal coordinates, which requires no analytical pre-derivation, making its creation self-contained. This process is an extension of the XY3 program \cite{05YuCaJe}. Although this procedure allows TROVE to simulate the nuclear motion for an arbitrary molecule, it does not calculate an exact kinetic energy operator. As such, the accuracy of the rotation-vibration energies depends on the expansion order of the kinetic energy operator and its associated level of convergence. Similarly, the potential energy operator is expressed as a Taylor-type expansion, and its numerical derivation can lead to an accumulation of round-off errors.
For a more detailed description of the TROVE functionality, including its treatment of the kinetic and potential energy operators, the reader is referred elsewhere \cite{07YuThJe}.

\subsection{Convergence}
To construct the ro-vibrational basis set, the contraction scheme described perviously \cite{jt500,jt503,11YaYuJe.H2CO} was followed. This method employs 1-D basis functions $\phi_{i}(\xi_{i}\lnr)$ ($i=1\ldots 6$), where $\xi_{1}\lnr$, $\xi_{2}\lnr$, and $\xi_{3}\lnr$ are the linearized versions of the three stretching coordinates $\Delta r_i$, $i=1,2,3$ ($\Delta r_i = {r_{\rm PH}}_i - r_{\rm e}$), and $\xi_4\lnr$, $\xi_5\lnr$, and $\xi_6\lnr$ are the linearized versions of the three bending coordinates $\Delta \alpha_i$, $i=1,2,3$ associated with the variation of the bond angles
$\alpha_{23}$, $\alpha_{13}$, and $\alpha_{12}$, respectively. Here $r_{\rm e}$ and $\alpha_{\rm e}$ are, respectively, the equilibrium values for the bond length and the inter bond angle facing the bond length.

The primitive basis functions for each mode $\nu_{i}$, $i=1\ldots 6$, are generated by solving the corresponding Schr\"{o}dinger equation with the Numerov-Cooley method~\cite{numerov,Cooley}.
In order to control the size of the basis sets at all contraction steps the polyad-truncation scheme \cite{07YuThJe} is used, based on the PH\3\ polyad number defined by
\begin{equation}\label{e:polyad-P}
    P =  2 (s_1 + s_2 +s_3) + b_{1} + b_{2} + b_{3} \le P_{\rm max}
\end{equation}
where $s_i$ and $b_i$  are the quantum numbers associated with the basis functions, $\phi_{s_i}$ and $\phi_{b_i}$, for the stretching modes and the bending modes, respectively. The basis set is formed only by those basis functions $\phi_s$, $\phi_b$ for which $P$ $\leq$ $P_{\rm max}$.

The primitive basis functions obtained at this step are then used to solve two 3-D Schr\"{o}dinger equations for each class of modes, stretching and bending, separately.  The corresponding basis sets are formed as direct products of the primitive basis functions satisfying $2 (s_1 + s_2 +s_3)\le{P_{max}}$ (stretching) and $b_{1} + b_{2} + b_{3}\le{P_{max}}$ (bending) in agreement with Eq.~\eqref{e:polyad-P}, so the maximal vibrational excitation is taken to be $P_{\rm max}/2$ and $P_{\rm max}$ for each of the stretching and bending modes, respectively.
The resulting two sets of eigenfunctions $\Phi_{m}^{\rm str}$ and $\Phi_{n}^{\rm bend}$ are then classified using $C_{ 3v}$(M) symmetry by analysing their transformational properties and assigned with the normal mode quantum numbers $n_1,n_3^{L_3}$ (stretching modes) and $n_2,n_4^{L_4}$ (bending modes) for future ease of line assignment and comparison to the experimental spectroscopic data. To this end a correlation between the primitive and normal mode quantum numbers is needed to be constructed. At the initial $J=0$ stage, this is a straightforward process due to the relatively small number of solutions and simple polyad structure of PH\3\ energies: only approximately $700$ sets of vibrational quantum numbers have to be translated for all energy levels ($J\le 31$) to be represented by both sets of quantum numbers.

The reassignment from local mode quantum numbers $\nu_i$ $(i=1\ldots 6)$ to normal mode quantum numbers $n_i$ $(i=1\ldots 4)$, $L_3 = |l_3|$ and $L_4 = |l_4|$  was performed by applying the following rules :
\begin{eqnarray}
   s_{1} + s_{2} + s_{3} &=& n_{1} + n_{3} ,  \nonumber \\
   b_{1} + b_{2} + b_{3} &=& n_{2} + n_{4} ,  \nonumber \\
   l_{3} &=& -n_{3}, -n_{3} + 2, ..., n_{3} - 2, n_{3} , \nonumber \\
   l_{4} &=& -n_{4}, -n_{4} + 2, ..., n_{4} - 2, n_{4}. \nonumber
\end{eqnarray}

It was also assumed that the symmetric modes $\nu_1$ and $\nu_2$ have lower energies than their asymmetric equivalents $\nu_3$ and $\nu_4$, respectively, and also that the vibrational energy grows when $l_3$ or $l_4$ increase.

At the next contraction step, a 6-D vibrational basis is formed as a direct product of the stretching and bending functions $\Phi_{m}^{\rm str}$ and $\Phi_{n}^{\rm bend}$, contracted using the (normal mode) version of the polyad number truncation given by Eq.~\eqref{e:polyad-P}:

\begin{equation}\label{e:polyad-normal}
    P =  2 (n_1 + n_3) + n_{2} + n_{4} \le P_{\rm max}.
\end{equation}

The 6-D functions $\Phi_{m}^{\rm str}$, $\Phi_{n}^{\rm bend}$ are then symmetrized by the standard reduction technique \cite{BJ}. At the next stage the vibrational ($J=0$) Schr\"{o}dinger equation is solved variationally by diagonalizing the ($J=0$) Hamiltonian matrix on this 6-D vibrational basis. Using the  $P_{\rm max}$ polyad restriction from \eqref{e:polyad-normal} 1455, 1125, and 2571 basis functions are obtained for the $A_1$, $A_2$ and $E$ symmetries, respectively.

At the last contraction step, the resulting eigenfunctions of the 6-D $J=0$ problem are used to form our final ro-vibrational basis functions in the so-called $J=0$ representation \cite{jt500} through a direct product with the symmetrized rigid rotor wave functions $\vert J,K,\tau \rangle$, where $\tau = 0,1$ indicates the rotational parity $(-1)^{\tau}$ as defined in Ref.\cite{05YuCaJe}. In the $J=0$ representation building the Hamiltonian matrix is straightforward as the vibrational part of the matrix is diagonal.

The ro-vibrational eigenfunctions obtained can be naturally assigned using a scheme based on the largest contribution of the basis set expansion. 
The chosen quantum numbers are the vibrational normal mode quantum numbers $n_1,n_2,n_3^{L_3}$ and $n_4^{L_4}$, the rotational quantum numbers $J,K$, 
the vibrational symmetry $\Gamma_{\rm rot}$, the vibrational symmetry $\Gamma_{\rm vib}$, and the total symmetry $\Gamma_{\rm tot}$, 
where $\Gamma_{\rm rot/vib/tot}=A_1,A_2,E$. The projection of the total vibrational 
 angular momentum $L$ is also included to reduce ambiguity in the description of the energy levels, with the following rules: 
$L$ must be a multiple of $3$ if $\Gamma_{\rm vib}=A_1,A_2$ and cannot be a multiple of 3 if $\Gamma_{\rm vib}=E$.  
By convention, the lower energy  value was assigned to $|L_3-L_4|$. However, at higher energies, TROVE does not necessarily assign unique quantum labels to every
state. In this case we have chosen to assign all states to $|L_3-L_4|$.

The accuracy with which high ro-vibrational states can be computed depends heavily on the size of the Hamiltonian matrix. To ensure that the calculation of the energy levels was sufficiently complete, convergence tests were performed to facilitate the choice of the polyad number ($P_{\rm max}$), ranging from $P=4$ to
$P=18$. A selection of the results from the convergence tests is given in table \ref{polyads}.
\begin{center}
\begin{table}[!h]
\small \tabcolsep=10pt
\renewcommand{\arraystretch}{1.3}

\caption[]{$J=0$ energy eigenvalue convergence with polyad number, $P_{\rm max}$,  used to generate the basis set, where $P(n)= (P_{max}=18) - (P_{max}=n)$.\label{polyads}} 
\begin{tabular}{{l}{r}{r}{r}{r}{r}r}
\hline\hline
{\textbf{Band}} &{\textbf{P$_{\rm max}=$18}} &  {\textbf{P(16)}} & {\textbf{P(14)}} & {\textbf{P(12)}} & {\textbf{P(10)}} &{\textbf{P(8)}}\\
\hline
$\nu_{0}$	&	0.000	&	0.000	&	0.000	&	0.000	&	0.000	&	0.000\\
$\nu_{2}$	&	992.732	&	-0.010	&	-0.033	&	-0.090	&	-0.243	&	-0.714\\
$2\nu_{2}$	&	1967.427	&	-0.133	&	-0.393	&	-0.939	&	-2.197	&	-5.321\\
$2\nu_{4}$	&	2222.257	&	-0.009	&	-0.033	&	-0.095	&	-0.283	&	-0.897\\
$\nu_{1}$	&	2324.077	&	-0.003	&	-0.011	&	-0.036	&	-0.128	&	-0.436\\
$3\nu_{2}$	&	2911.866	&	-1.325	&	-3.472	&	-7.230	&	-14.335	&	-29.377\\
$\nu_{2}+2\nu_{4}$	&	3204.920	&	-0.115	&	-0.361	&	-0.941	&	-2.467	&	-7.220\\
$\nu_{1}+\nu_{2}$	&	3322.008	&	-0.038	&	-0.132	&	-0.384	&	-1.135	&	-3.820\\
$3\nu_{4}$	&	3341.959	&	-0.028	&	-0.098	&	-0.295	&	-0.931	&	-3.434\\
$\nu_{3}+\nu_{4}$	&	3439.304	&	-0.009	&	-0.034	&	-0.117	&	-0.437	&	-1.843\\
$4\nu_{2}$     &	3803.487	&	-9.750	&	-22.284	&	-39.911	&	-66.534	&	-108.521\\
$2\nu_{2}+2\nu_{4}$	&	4159.298	&	-1.025	&	-2.839	&	-6.309	&	-13.381	&	-28.564\\
$\nu_{1}+2\nu_{2}$	&	4303.413	&	-0.352	&	-1.074	&	-2.740	&	-6.742	&	-16.622\\
$\nu_{2}+3\nu_{4}$	&	4318.966	&	-0.257	&	-0.786	&	-1.981	&	-4.906	&	-12.643\\
$\nu_{2}+\nu_{3}+\nu_{4}$	&	4410.959	&	-0.131	&	-0.436	&	-1.196	&	-3.221	&	-9.642\\
$4\nu_{4}$	&	4422.841	&	-0.098	&	-0.325	&	-0.919	&	-2.707	&	-7.461\\

\hline\hline

\end{tabular}
\end{table}
\end{center}

Ideally, the polyad number $P_{\rm max}$ would be chosen to give full convergence, which, as can be seen from table \ref{polyads}, is not achieved even for the highest polyad configuration, $P_{\rm max}=18$. However, although the calculations should improve as the polyad number increases, since the `spectroscopic' PES (see Section {\bf{3.2}}) used was generated with $P_{\rm max} =14$, the energy levels move away from the observed values when a higher polyad configuration is used. The PES used in this work relies on a self consistency of parameters, and as such it is only an effective potential for a specific set of parameters, so $P_{\rm max} =14$ was chosen.

Convergence tests were performed to test the expansions of the kinetic and potential energy operator and, although accuracy did improve considerably as the expansions grew, so did the computational cost involved. The convergence test presented in Table \ref{polyads} was done using expansions of the kinetic energy operator and potential energy function both to  fourth order, which accounts for the significant discrepancies between the computed band origins and the corresponding experimental values. In the present work, the expansions of the kinetic energy operator and potential energy function are truncated at the 6$^{th}$ and 8$^{th}$ order, respectively, which leads to much more accurate results, as can be seen below. Higher order expansions than these would have made the project computationally prohibitive.

\subsection{Potential Energy Surface}
The PES of phosphine used here is a refinement of the {\it{ab initio}} (CCSD(T)/aug-cc-pV(Q+d)Z) PES \cite{08OvThYu2}, done by performing a least square fit to available experimental ro-vibrational energies with $J=0, 1, 2, 4$ and $10$.
 These refinements followed the fitting procedure introduced in Ref. \cite{jt503}.
The refinement of the PES is represented as a correction $\Delta V$ to the \abinitio\ PES, $V_0$. The ro-vibrational $J=0,1,2,4,10$ eigenfunctions  of the `unperturbed' Hamiltonian $H_0 = T + V_0$ are used as basis functions when iteratively solving the set of Sch\"{o}dinger equations for the ro-vibrational behaviour of the molecule to minimise the associated functional in the least-squares fitting. With the final refined PES, a total root-mean-squares (rms) error of 0.03 \cm\ was obtained for the fitting energy set.
Table \ref{tm} shows the change in the band centre values between the pre- and post-refinement PES.

As already mentioned, the kinetic energy operator is not calculated exactly and the basis set is artificially limited. Thus the refined PES must be considered an effective PES; it only gives the accurate results presented in Table \ref{tm} when used with TROVE and the parameters described above. The refined PES used here is given in the supplementary material to this paper in the form of a Fortran program.

The PES is a major source of error in the line list creation process, and as such constant updates and refinements are necessary. For example, although the current PES is appropriate for creating an accurate room temperature line list with wavenumbers up to 8000 cm$^{-1}$ and $J\leq31$, it would not necessarily remain so for a line list with parameters exceeding this.
For example, the PES fails to accurately describe the potential of phosphine at very high excitations, showing some artificial minima in the region of dissociation. Consequently, this PES is not suitable for use in dynamical simulations without adjustment.

\subsection{Dipole moment surface and transition intensities}

An existing six-dimensional {\it{ab initio}} electric dipole moment (CCSD(T)/aug-cc-pVTZ) \cite{06YuCaTh} is used to obtain the Einstein coefficients and transition intensities. This was calculated on a large grid of $10080$ molecular geometries. The  procedure used to compute absolute intensities mirrored that used for ammonia, and is described in detail in Yurchenko et al \cite{jt466}.
A number of vibrational transitional moments were calculated to help characterize the quality of the DMS.

Table \ref{tm} compares the empirical vibrational transition moment with values computed by TROVE, both in the present work and in the previous studies  \cite{06YuCaTh}. It is clear that the theoretical band intensities are in good agreement with experiment and, although the DMS used here is unchanged from that used in Ref. \cite{06YuCaTh}, the use of our more accurate PES (and hence wavefunctions) has meant that our new results have generally reduced the error in the calculations of the transition moments, from an average deviation of $22.5\%$ to $10\%$.

\begin{center}
\begin{table}[h!]
\footnotesize \tabcolsep=10pt
\renewcommand{\arraystretch}{1.4}
\caption[]{Calculated band centres and their respective transition moments (deviation from experimental values shown as a percentage). Uncertainties of the experimental (Obs)  transition moments are given in parentheses (in units of the last digit quoted) where available.\label{tm}}
\begin{tabular}{{c}{r}{r}{r}{c}{l}{l}{l}}
\hline\hline
&\multicolumn{3}{c}{Band Centres (cm$^{-1}$)} && \multicolumn{3}{c}{Transitions Moments (D$^{2}$)}\\
\cline{2-4}\cline{6-8}
{\bf{Band}} & {\bf{Obs}} & {\textbf{Calc '06  \cite{06YuCaTh}}} & {\textbf{Present}} &&  {\bf{Obs}} &  {\textbf{Calc '06 (\%) \cite{06YuCaTh}}} & {\textbf{Present (\%)}} \\
\hline
$\nu_{0}$	&0.000	    & 0.000    &	0.000 &&0.57395(30) &	 0.583(1.6\%) & 0.585(2.0\%)\\
$\nu_{2}$	&992.130	& 992.500  & 992.152  &&0.08251(5) &	0.085(2.5\%)&0.084(2.2\%)\\
$\nu_{4}$	&1118.310	& 1117.870 & 1118.322&&0.08626(5) & 0.087(0.3\%)&0.089(3.7\%)\\
$2\nu_{2}$	&1972.550	& 1972.820 & 1972.590&&0.00299(5) & 0.003(9.7\%)&0.004(21.4\%)\\
$\nu_{2}+\nu_{4}$	&2108.150	& 2107.170 & 2108.169&&0.01102(6) & 0.009(15.6\%)&0.014(25.3\%)\\
$2\nu_{4}$	&2226.830	& 2227.860 & 2226.835&&0.0176(2) & 0.006(68.8\%)&0.018(1.5\%)\\
$2\nu_{4}$	&2234.930	& 2234.570 & 2234.940&&0.0176(2) & 0.002(91.5\%)&0.013(23.5\%)\\
$\nu_{1}$	&2321.120	& 2322.040 & 2321.142&&0.0690 & 0.073(6.1\%)&0.072(4.3\%)\\
$\nu_{3}$	&2326.870	& 2329.180 & 2326.888&&0.130 & 0.139(6.8\%)&0.138(6.1\%) \\
\hline\hline

\end{tabular}
\end{table}
\end{center}

\subsection{Empirical adjustment of the vibrational band centres}

Although using a `spectroscopic' PES improves the values of the energy levels, the TROVE calculations do not completely reproduce the observed transition frequencies. To correct this, an empirical approach was adopted, where an artificial frequency shift is added to calculated band origins as given by the $J=0$ energies; this procedure is described as an empirical basis set correction scheme (EBSC) \cite{jt466}. It leads to a rotational energy structure in much better agreement with experimental results for the remaining J values of the band. As a test, the subsequent $J=1,2$ and $4$ values were compared to those from experiments which, together with $J=0$ were also the values of $J$ used to refine the PES. While using this method, care must be taken to only select reliable experimental data, since data of limited accuracy can be a source of error which TROVE cannot compensate for.

Only 11 band centres were manipulated, and several iterative shifts were attempted. The final improvements to the standard deviation, $\sigma$, of the whole band from experiment are displayed on table \ref{BandCentres}. The original average $\sigma$ value for the set of bands was $0.031$ cm$^{-1}$ which was reduced to $0.02$ cm$^{-1}$ after this final adjustment.\\

\begin{table}[!h]
\small \tabcolsep=11pt
\renewcommand{\arraystretch}{1.4}

 \caption{Observed band origins (Obs) and standard deviation
with which TROVE reproduces the terms within each band, $\sigma$,
before and after replacement of the band origins.
Observed data is from HITRAN2008 \cite{jt453}.}

\begin{tabular}{cccc}
\hline\hline
\multicolumn{1}{c}{\textbf{Band}} &\multicolumn{1}{r}{\textbf{Obs}} & \multicolumn{1}{c}{\textbf{Original $\sigma$}} & \multicolumn{1}{c}{\textbf{New $\sigma$ }}  \\
\hline
$\nu_{2}$	&$992.135$ & 0.030 & 0.020 \\
$\nu_{4}$&$1118.307$ & 0.012 & 0.005 \\
$2\nu_{2}$	&$1972.571$ & 0.011& 0.007 \\
$\nu_{2}+\nu_{4}$	&$2108.152$ & 0.045 & 0.034 \\
$2\nu_{4}$	&$2226.835$ & 0.029 & 0.010  \\
$2\nu_{4}$	&$2234.920$ & 0.032 & 0.014  \\
$\nu_{1}$	&$2321.121$ & 0.028 & 0.012  \\
$\nu_{3}$	&$2326.8667$ & 0.023 & 0.013  \\
$3\nu_{2}$	&$2940.767$ & 0.063 & 0.046  \\
$\nu_{2}+2\nu_{4}$	&$3214.936$ & 0.035 &0.024 \\
$\nu_{3}+\nu_{4}$	&$3440.259$ & 0.037 & 0.034 \\
\hline\hline

\label{BandCentres}
\end{tabular}
\end{table}

\newpage

\section{Results}

\begin{figure}[htp!] \begin{center}
\includegraphics[scale=0.45]{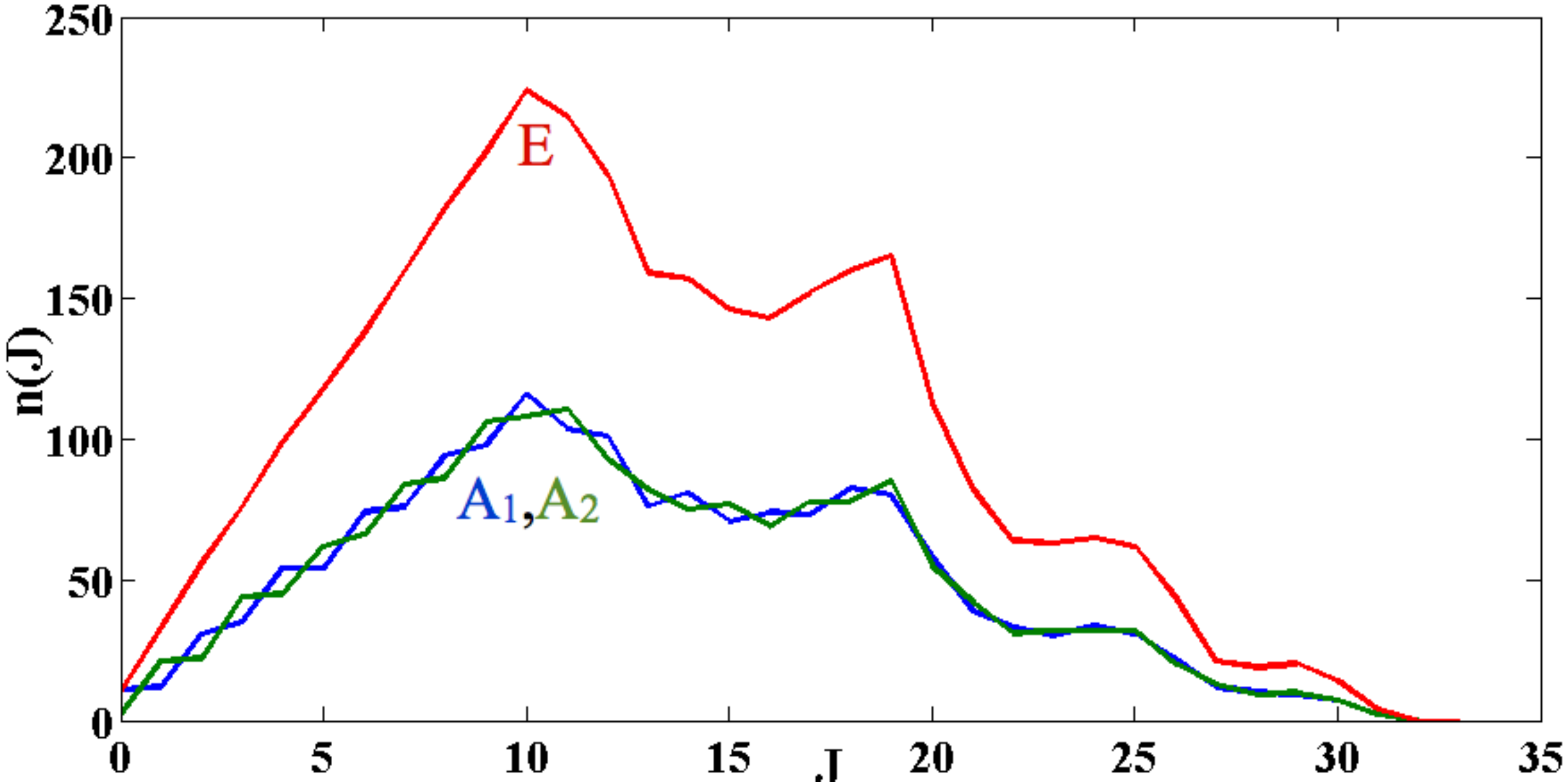}
\caption{ \label{eigenvalues} Total number of energy eigenvalues under $4000$ cm$^{-1}$ for $J=0,1,...,33$. The blue, green and red lines represent, respectively $A_{1}$, $A_{2}$ and $E$ symmetry eigenvalues.}
\end{center}  \end{figure}

For the room temperature line list presented here the following thresholds were selected.
The chosen ranges of energy eigenvalues are $4000$ cm$^{-1}$ for the highest lower energy and $12000$ cm$^{-1}$ for the highest upper energy. These choices allow for a range of transitions of $0$ to $8000$ cm$^{-1}$ at temperatures up to $300$ K. The lower energy threshold of $4000$~\cm\ defines the highest $J$ that had to be taken into account, as there were many more energy values for each symmetry and each $J$ than those within the selected range. By $J=31$ only $0.0025\%$ of all energy eigenvalues (a total of 8 levels, across all symmetries) were below $4000$ cm$^{-1}$ and by $J=32$ there were none.
Figure \ref{eigenvalues} shows the variation in number of useful (within range) eigenvalues as $J$ increases. 

Summing over all the energy levels calculated here gives a partition function equal to $3249.5$ at 296 K. This is $0.028\%$ higher than the latest published HITRAN value of $3248.6$ \cite{06SiJaRo}. Further work on the partition function and other thermodynamic properties of both phosphine and ammonia can be found in \cite{13SoHeYu}.\\

For the purposes of comparison with the existing experimental data, only transitions above a minimum intensity were included. In the region $0 - 300$ cm$^{-1}$, only transitions stronger than $\geq{10^{-13}}$ cm/mol  (or $1.7\times10^{-37}$ cm/molecule) were considered, while for transitions with wavenumbers $\geq300$ cm$^{-1}$ this threshold was reduced to ${10^{-5}}$ cm/mol ($1.7\times10^{-29}$ cm/molecule). These are approximately one order of magnitude more sensitive than the weakest lines in both the HITRAN 2008 \cite{jt453} and CDMS  \cite{05MuScSt} databases. With this intensity cut-off, only 5 488 177 transitions were selected from the total of 137 255 400 computed lines.

\begin{figure}[htp] \begin{center}
\includegraphics[scale=0.45]{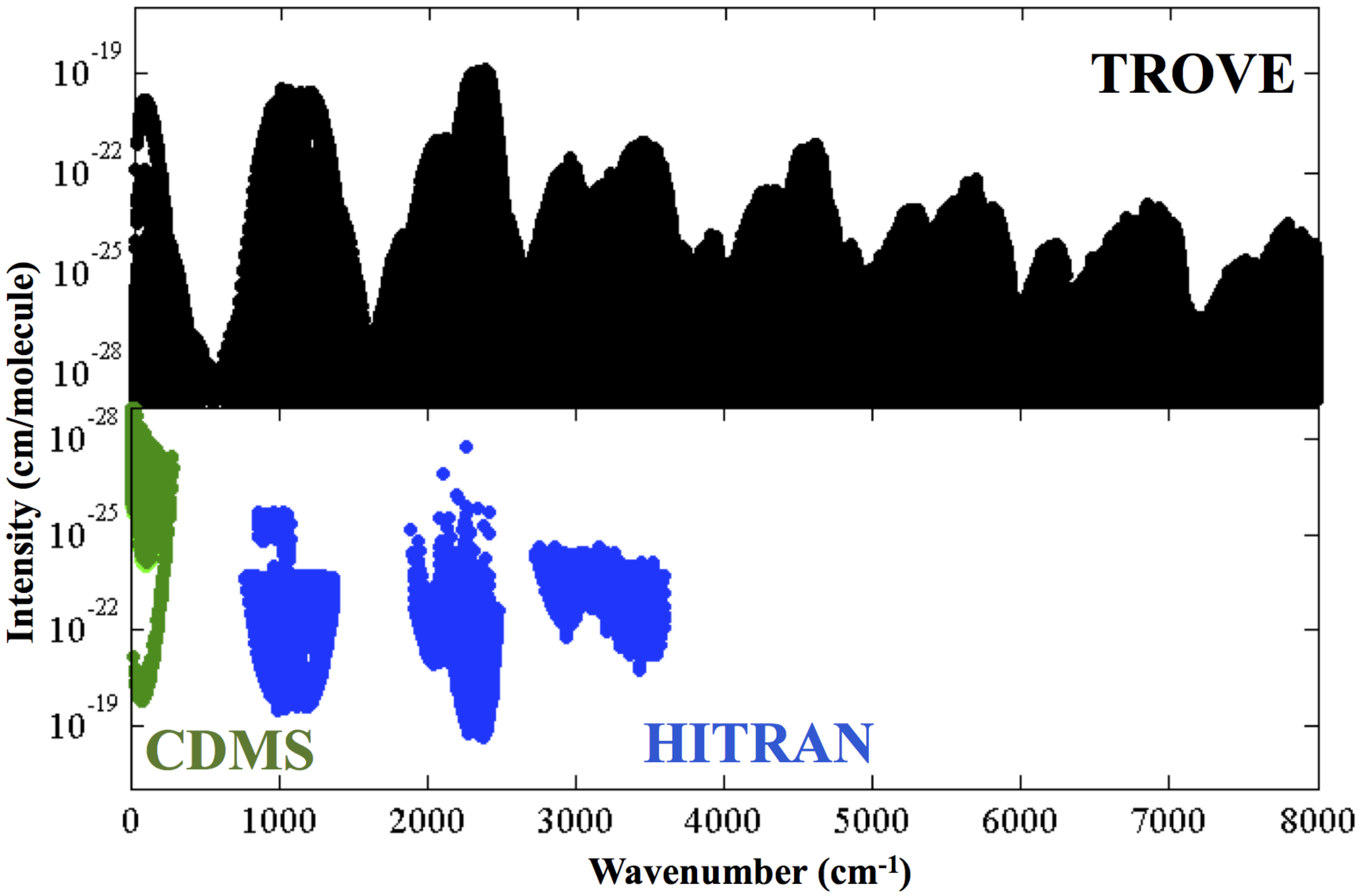}
\caption{\label{overview}Comparison between simulated absorption spectrum created by TROVE and those from the JPL \cite{98PiPoCo} and HITRAN \cite{jt453} databases, as a function of wavenumber.}
\end{center}
\end{figure}

Figure \ref{overview} shows an overview of our simulation compared to CDMS \cite{05MuScSt}  and HITRAN \cite{jt453} databases. HITRAN lacks any pure rotational transitions and, although it captures most lines stronger than $5 \times 10^{-25}$ cm/molecule, it is very incomplete below this value.

All the information on the $5.6$ million rotation-vibration energy levels and the associated list of the $137$ million transitions can be 
found online at {\it{www.exomol.com}}, in the ExoMol format \cite{jt548}. It is possible to use this line list to generate synthetic spectra as a function of temperature. However, such spectra will become increasingly incomplete as the temperature goes above $300$ K.
Additionally, the temperature-dependent molecular absorption cross sections \cite{jt542} for phosphine are also available online.
A HITRAN format line list is included as supplementary material to this paper. An intensity cut-off of $10^{-31}$ cm/molecule is used for the rotational part of the spectrum ($0-500$ cm$^{-1}$) and $10^{-28}$ cm/molecule for the remainder of the spectrum.
Additionally, the refined potential energy surface used here is provided as a fortran program.


Data for the rotational spectrum was obtained from the CDMS database \cite{05MuScSt}, although the strongest of these lines can also be found in the JPL database \cite{98PiPoCo,10PeMuPi}.  The documentation for the CDMS data can be found online and cites its sources as Cazzoli and Puzzarini \cite{06CaPuXX}, Belov et al \cite{81BeBuGe}, Fusina and Carlotti \cite{88FuCaXX}, Davis, Newman, Wofsy, and Klemperer \cite{71DaNeWo}, Chu and Oka \cite{74ChOkXX},  Helms and Gordy, \cite{77HeGoXX}, and Belov, Burenin, Polyanski, and Shapin \cite{81BeBuPo}. Further information about these papers can be found in Table \ref{sources}. The CDMS data set contains 2131 transition lines in the region 0 - 300 cm$^{-1}$ with a maximum $J$ quantum number of 34.

To adequately compare the lines calculated here to those in CDMS, lower and upper energy levels were matched between datasets and the difference between the resulting transitions analysed. Due to the ambiguity (between A$_{1}$ and A$_{2}$) in many of the energy levels with A symmetries, only the E symmetry transitions were matched with sufficient confidence for an accurate analysis. This sample can be expected to give a representative rms for the remaining transitions. An algorithm created with MATLAB matched theoretical lines produced by TROVE with equivalent ones in the CDMS dataset. These matches deviated from those in CDMS with a rms value of 0.076 cm$^{-1}$. Problems with ambiguous or incorrect labelling of energy levels skew the rms deviation and its true value is expected to be significantly lower. In fact, when the top $1\%$ worst matches are removed from the comparison, the rms deviation lowers to 0.05 cm$^{-1}$.

\begin{figure}[htp] \begin{center}
\includegraphics[scale=0.35]{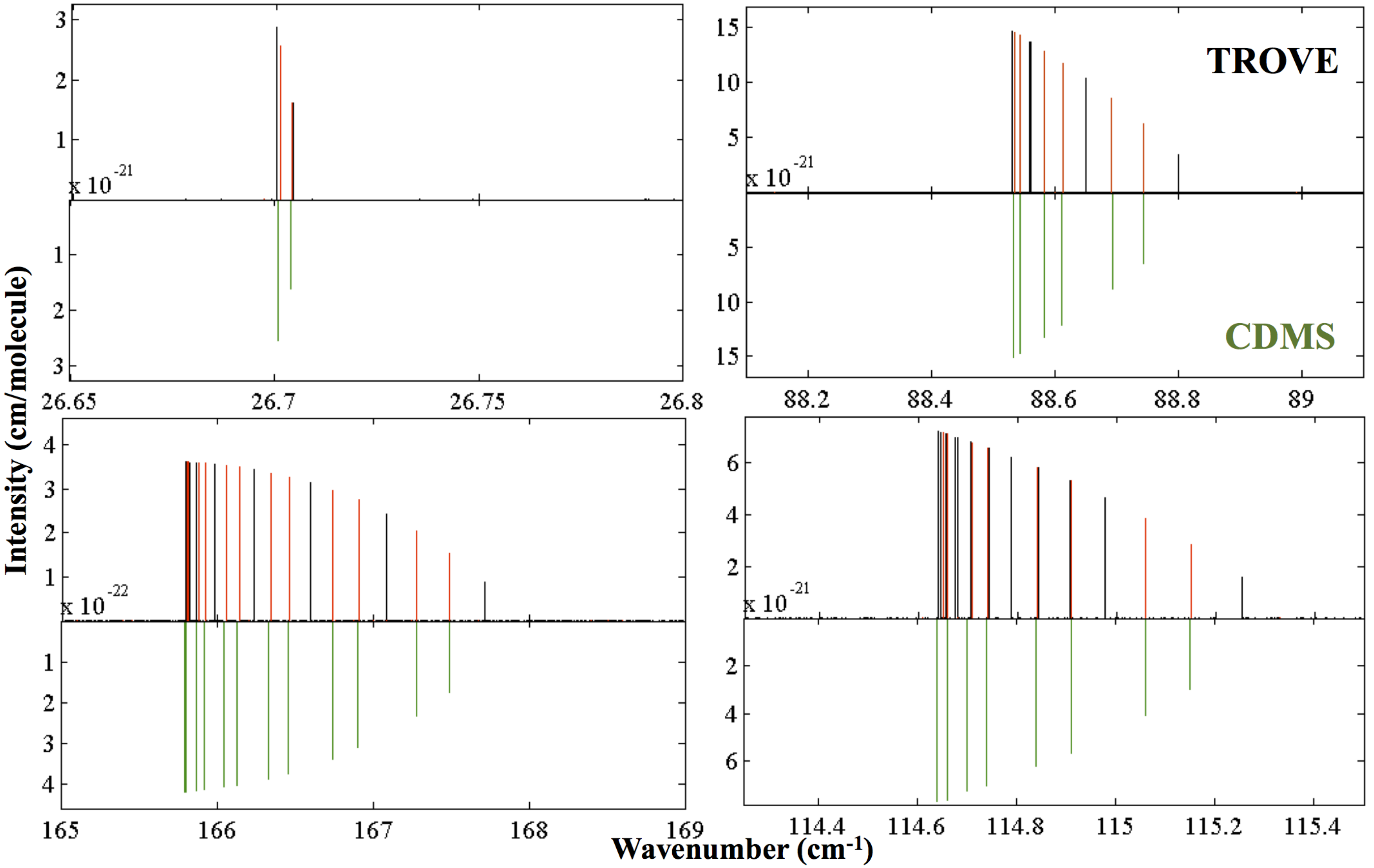}
\caption{\label{jplzoom}CDMS (lower) versus TROVE (upper). The matched TROVE transitions are highlighted in red.}
\end{center} \end{figure}
Figure \ref{jplzoom} shows close-ups of the rotational spectrum, where a line by line comparison can be made with the CDMS data mirroring TROVE's theoretical predictions, and those transitions that could be matched highlighted in red.

 
The most recent HITRAN data \cite{jt453} for phosphine contains 20~099 transitions in the region 770 - 3601 cm$^{-1}$. Of these, only 16~008 are assigned and of these 2011 have no upper vibrational quantum numbers, possibly due to vibrational mixing. There seems to be no consistent description of symmetry and many transitions appear either mislabelled or misassigned.

The energy levels of the assigned lines were compared to those calculated in TROVE and this information was used to match lines between experiment and theory. Similarly to the CDMS dataset, ambiguity between energy levels with A symmetry meant that only E symmetry transitions were considered in the comparison. Again, this sample can be expected to give representative rms for each band. Assigned E symmetry transitions correspond to 7838 of the total HITRAN lines. 13 of these transitions had ambiguous or incorrect K values, so only 7825 E transitions were matched. These were found to have a rms deviation from their experimental equivalent of 0.23 cm$^{-1}$. The wavenumbers of a few transitions involving energy levels which are defined by a single transition in the HITRAN database disagree significantly with that of those calculated with TROVE. These skew the rms deviation and again, its true value is expected to be significantly lower. When the top $1\%$ worst matches are removed from the comparison, the rms deviation lowers to 0.19 cm$^{-1}$.

\begin{figure}[htp!] \begin{center}
\includegraphics[scale=0.35]{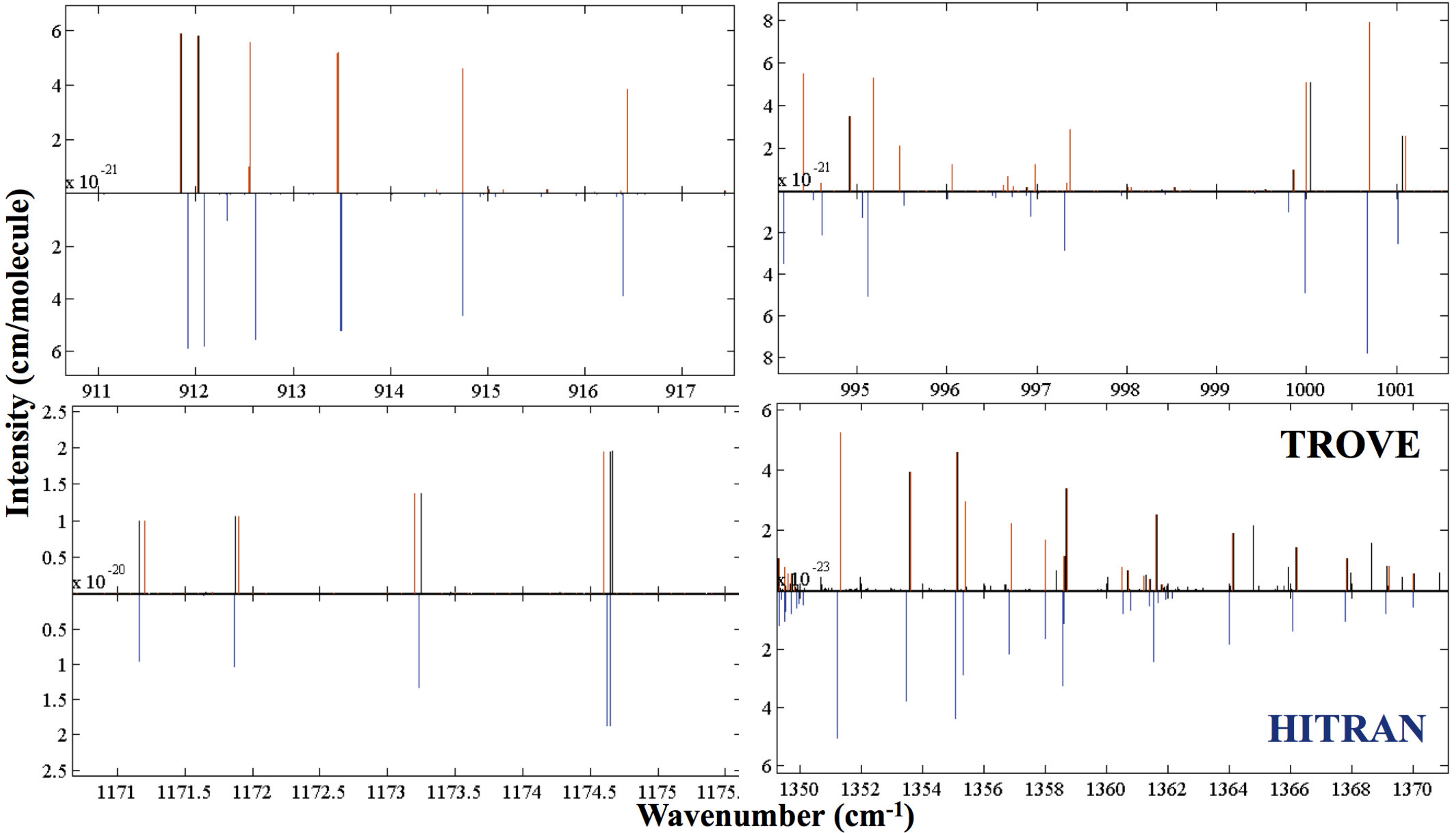}
\caption{\label{p1}HITRAN (lower) versus TROVE (upper), with matched transitions highlighted in red, for the region $P=1$.}
\end{center} \end{figure}

The first region (Polyad number = 1) is located between 770 -- 1372 cm$^{-1}$ and is dominated by the fundamental bending bands $\nu_{2}$ and $\nu_{4}$ and the ''hot" band $2\nu_{2}-\nu_{2}$. TROVE's rms deviation from HITRAN in this region is 0.23 cm$^{-1}$. Its rms is 0.37 cm$^{-1}$ for the $\nu_{2}$ band, 0.11 cm$^{-1}$ for the $\nu_{4}$ band and 0.11 cm$^{-1}$ for the $2\nu_{2}-\nu_{2}$ band. HITRAN references Brown, Sams and Kleiner \cite{02BrSaKl} as the source for this region.
Figure \ref{p1} shows close-ups of a selection of representative sub-regions within the $P=1$ region.\\

\begin{figure}[htp!] \begin{center}
\includegraphics[scale=0.35]{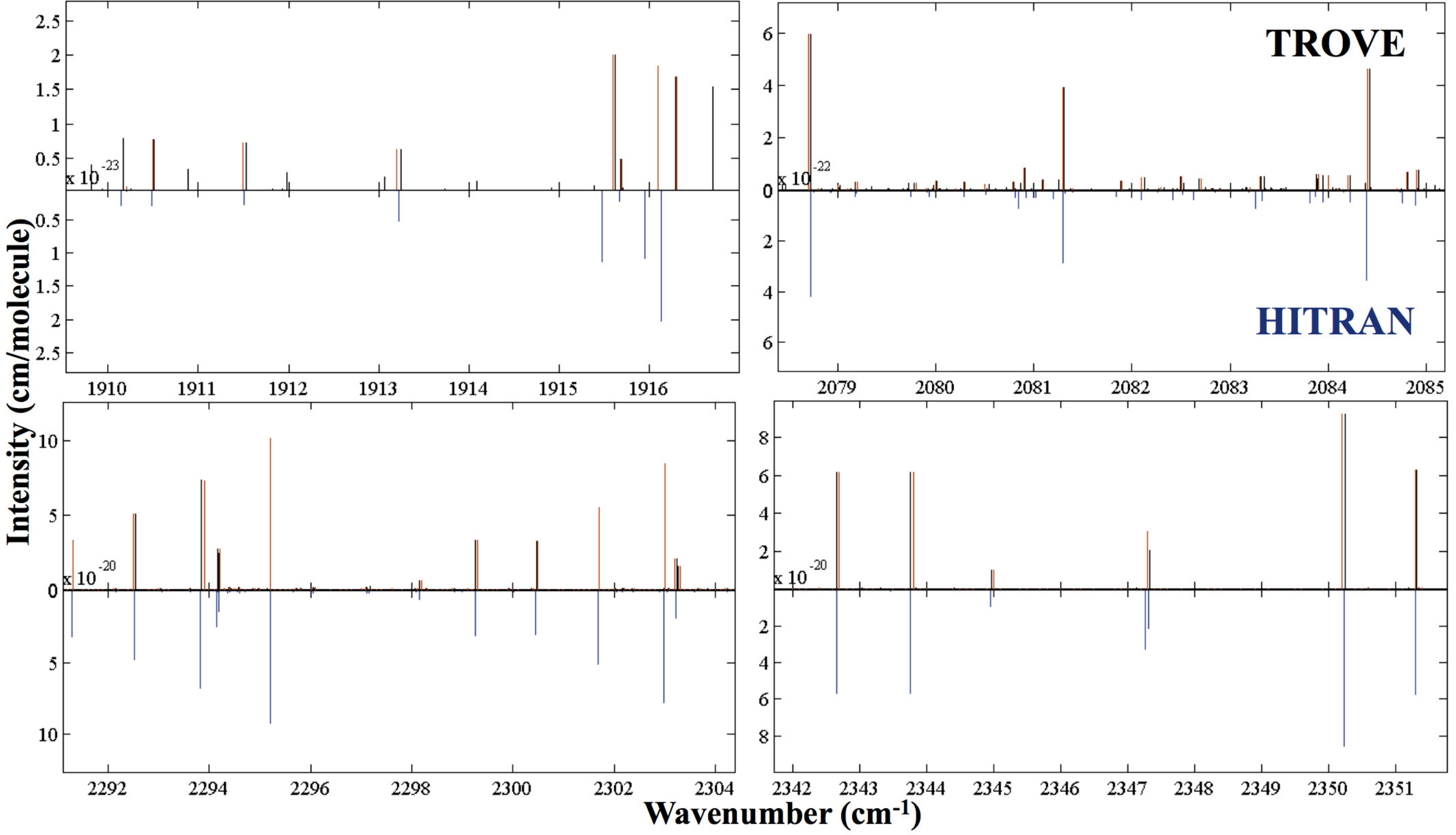}
\caption{\label{p2}HITRAN (lower) versus TROVE (upper), with matched transitions highlighted in red, for the region $P=2$.}
\end{center} \end{figure}

The second region is located between 1885 -- 2478 cm$^{-1}$ and is dominated by the fundamental stretching bands $\nu_{1}$ and $\nu_{3}$, the $2\nu_{2}$ and $2\nu_{4}$ fundamental overtones and the combination band $\nu_{2}+\nu_{4}$. TROVE's rms deviation from HITRAN in this region is 0.20 cm$^{-1}$, with 0.28 cm$^{-1}$ for the $\nu_{1}$ band, 0.22 cm$^{-1}$ for $\nu_{3}$, 0.05 cm$^{-1}$ for $2\nu_{2}$, 0.15 cm$^{-1}$ for $2\nu_{4}$ and 0.08 cm$^{-1}$ for $\nu_{2}+\nu_{4}$. HITRAN references Tarrago et al \cite{92TaLaLe} as the source for this region.
Figure \ref{p2} shows close-ups of a selection of representative sub-regions within the $P=2$ region.\\

\begin{figure}[htp!] \begin{center}
\includegraphics[scale=0.35]{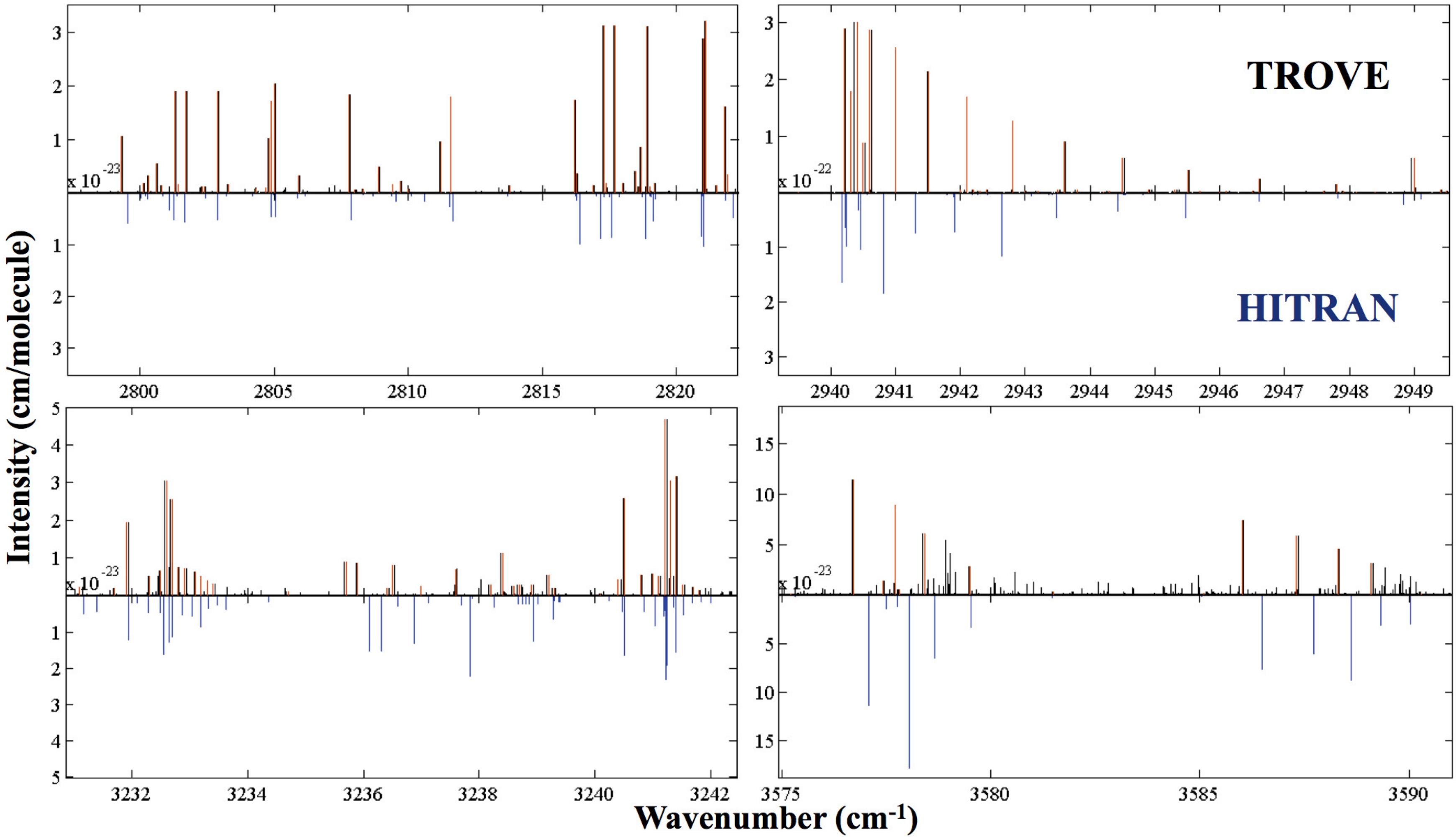}
\caption{\label{p3}HITRAN (lower) versus TROVE (upper), with matched transitions highlighted in red, for the region $P=3$.}
\end{center} \end{figure}

The third region is located between 2733 -- 3600 cm$^{-1}$ and is dominated by the$4\nu_{2}-\nu_{2}$ hot band, the $3\nu_{2}$ overtone and the $\nu_{1}+\nu_{2}$, $\nu_{1}+\nu_{4}$, $\nu_{2}+\nu_{3}$, $\nu_{3}+\nu_{4}$, $\nu_{2}+2\nu_{4}$ and $2\nu_{2}+\nu_{4}$ combination bands. TROVE's rms deviation from HITRAN in this region is 0.27 cm$^{-1}$, with 0.46 cm$^{-1}$ for $4\nu_{2}-\nu_{2}$, 0.37 cm$^{-1}$ for $3\nu_{2}$, 0.50 cm$^{-1}$ for $\nu_{1}+\nu_{2}$, 0.21 cm$^{-1}$ for $\nu_{1}+\nu_{4}$, 0.23 cm$^{-1}$ for $\nu_{2}+\nu_{3}$, 0.19 cm$^{-1}$ for $\nu_{3}+\nu_{4}$, 0.11 cm$^{-1}$ for $\nu_{2}+2\nu_{4}$ and 0.28 cm$^{-1}$ for $2\nu_{2}+\nu_{4}$. HITRAN references Butler et al \cite{06BuSaKl} as the source for this region.
Figure \ref{p3} shows close-ups of a selection of representative sub-regions within the $P=3$ region.\\


Nikitin et al \cite{09NiHoTy} computed 55~223 transitions in the range 700 -- 3500 cm$^{-1}$ (the HITRAN range \cite{jt453}). This work does not aim for completeness and even within the HITRAN regions appears to have significant omissions (e.g. maximum $J=20$). The work presented here improves on their root mean squares deviation from observed band centres of 1.4 cm$^{-1}$.

Experimental data (1768 lines) from Wang et al \cite{05WaChCh} in the 1950-2480 cm$^{-1}$ and 3280-3580 cm$^{-1}$ regions was compared to the theoretical lines created by TROVE.  The results presented here deviated from those of Wang et al's with an overall rms deviation of 0.17 cm$^{-1}$, for those lines that could be matched with confidence. The rms deviation for the first region was 0.11 cm$^{-1}$, compared to 0.20 cm$^{-1}$ for the equivalent HITRAN region and 0.21 cm$^{-1}$ for the second, compared to 0.27 cm$^{-1}$ for HITRAN.

\begin{figure}[htp] \begin{center}
\includegraphics[scale=0.4]{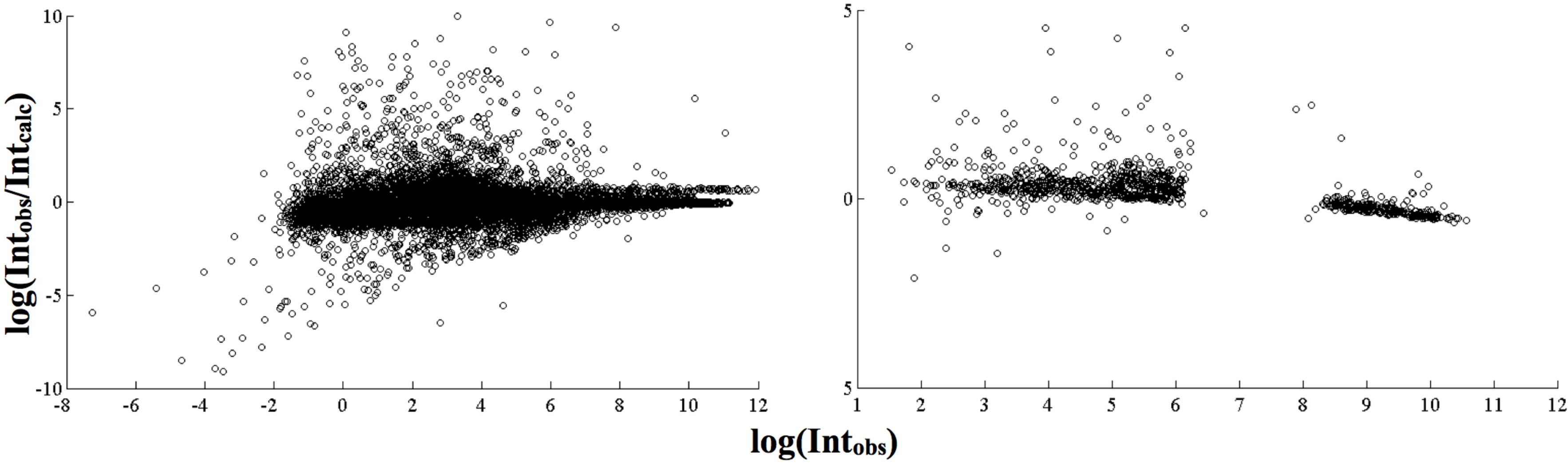}
\caption{\label{LogIntPlot} Plot of $log(I_{obs}/I_{calc})$ versus $log(I_{obs})$ for the HITRAN data \cite{jt453} (left) and the Wang data  \cite{05WaChCh}.}
\end{center} \end{figure}

 Additionally, our intensities agree significantly better with the Wang data than with HITRAN's data, as can be seen by the intensity plots in Figure \ref{LogIntPlot}, suggesting that Wang et al's intensities should be used in a future release of HITRAN.

The JPL data  \cite{10PeMuPi, 98PiPoCo} agrees very well with our calculations. It contains 729 transitions between $0 -188$  cm$^{-1}$ with J$\leq27$. Energy level comparison of the E symmetry transitions showed that the data presented here deviated from JPL with a rms of $0.066$ cm$^{-1}$. When the top $1\%$ worst matches are removed from the comparison, the rms deviation lowers to 0.031 cm$^{-1}$. Overall, the JPL transition wavenumbers agree more closely with those presented here than CDMS'. However, the hundred strongest transitions in the rotational spectrum from CDMS match ours much better ($0.003$ cm$^{-1}$ compared to JPL's $0.07$ cm$^{-1}$). Additionally, CDMS is considerably more complete and its intensities are in closer agreement with ours. This should at least partially be due to the fact that some of the K=3n transition doublets are very close so their intensities are combined in the JPL database, while the equivalent TROVE lines are left resolved. This difference in data handling leads to an apparent factor of two disagreement in the intensities.

\section{Discussion and Conclusion}

In this work a phosphine line list is produced with $137$ million transitions between $5.6$ million energy
 levels for ro-vibrational states up to $J_{max}=31$ and energies up to $8000$ cm$^{-1}$. This work replicates
 very well the observed phosphine spectra at room temperature, with a maximum rms deviation from CDMS of  0.076 cm$^{-1}$ 
for the rotational spectrum and of 0.23 cm$^{-1}$ from HITRAN. This is a valid line list for any phosphine analysis below 300 K, 
however our ultimate aim is to construct a hot line list capable of simulating observed spectra from astronomical bodies at higher temperatures, e.g. 2000 K. 
This line list  will complement the one already available for ammonia \cite{jt500}.
Work on this is currently in progress.

 The tunnelling effect present in the ammonia molecule is predicted to be found in phosphine \cite{81BeBuPo} but, due to its much higher barrier (12~300 cm$^{-1}$), yet to be observed. The value of splitting in various vibrational states as well as the intensity of the inversion-rotation, and inversion-rovibrational lines can be computed by adapting the procedure given here to work with D$_{3h}$(M) symmetry. Predictions for transitions which can be used to resolve the doublet splitting will be presented in future work.

\vspace{0.5in}

\section*{Acknowledgments}
This work is supported by ERC Advanced Investigator Project 267219. We would like to also thank Dermot Madden, Oleg Polyansky and Charles Leahy for their support and contribution.

\vspace{0.5in}

*To whom correspondence should be addressed. clara$\_$ss$@$star.ucl.ac.uk


\end{document}